\documentclass{sig-alternate}

\usepackage{color}
\usepackage{subfigure}
\usepackage{url}
\usepackage{multirow}
\usepackage{fancyvrb}
\fvset{frame=single,samepage=true,fontfamily=courier,numbers=left,numbersep=2pt,fontsize=\tiny}

\usepackage{listings}
\lstnewenvironment{example}[1][]{
    
    \lstset{#1}
}{}

\newcounter{codecounter}
\begin{document}

\title{HTTP Mailbox - Asynchronous RESTful Communication}

\numberofauthors{3}

\author{
\alignauthor
Sawood Alam\\
\affaddr{Dept. of Computer Science}\\
\affaddr{Old Dominion University}\\
\affaddr{Norfolk, Virginia, USA}\\
\email{salam@cs.odu.edu}
\alignauthor
Charles L. Cartledge\\
\affaddr{Dept. of Computer Science}\\
\affaddr{Old Dominion University}\\
\affaddr{Norfolk, Virginia, USA}\\
\email{ccartled@cs.odu.edu}
\alignauthor
Michael L. Nelson\\
\affaddr{Dept. of Computer Science}\\
\affaddr{Old Dominion University}\\
\affaddr{Norfolk, Virginia, USA}\\
\email{mln@cs.odu.edu}
}

\maketitle
\begin{abstract}

We describe HTTP Mailbox, a mechanism to enable RESTful HTTP communication in an asynchronous mode with a full range of HTTP methods otherwise unavailable to standard clients and servers. HTTP Mailbox allows for broadcast and multicast semantics via HTTP. We evaluate a reference implementation using ApacheBench (a server stress testing tool) demonstrating high throughput (on 1,000 concurrent requests) and a systemic error rate of 0.01\%. Finally, we demonstrate our HTTP Mailbox implementation in a human assisted web preservation application called ``Preserve Me!''.

\end{abstract}




\section{Introduction}
\label{intro}

Alice wanted to keep track of her tasks and maintain a to-do list. She found Bob's shared, hosted task manager service. She created her initial tasks list (Table~\ref{tab:tasks}) and started working on the highest priority task. Once it was finished, she wanted to mark that task as done. Hence, Alice made an HTTP PATCH request~\cite{rfc5789} to the specific task URI on Bob's server to  modify the task partially. Unfortunately the server was down so this communication failed. Alice tried again after some time when the server was up but got a \texttt{501 Not Implemented} HTTP Response~\cite{rfc2616} from the server. Alice talks to Bob regarding this issue and Bob replied that there is a non-RESTful way of doing this on the task server. They wished there was an extra layer of indirection to provide a RESTful interface to Bob's server, even if indirectly.

\begin{table}[h]
  \centering
  \caption{Alice's tasks}
  \label{tab:tasks}
  \begin{tabular}{|l|l|l|l|}
    \hline
    \textbf{ID} & \textbf{Description} & \textbf{Priority} & \textbf{Status}\\
    \hline
    1 & Write a paper. & HIGH & Pending\\
    \hline
    2 & Go on vacation. & LOW & Pending\\
    \hline
  \end{tabular}
\end{table}

HTTP Mailbox provides a layer of indirection. It allows sending any HTTP message (request or response), encapsulated in the message body to a URI relative to the HTTP Mailbox service using an HTTP POST request. Resulting messages can be retrieved by making an HTTP GET request to the HTTP Mailbox. Multiple HTTP messages to the same recipient can be pipelined in a single HTTP POST request. The HTTP Mailbox also provides multicast and broadcast messaging capabilities, an enhancement not possible using HTTP.

In past years, general web services used only the GET and POST methods of the HTTP protocol while several other HTTP methods like PUT, PATCH, and DELETE were rarely used. Until recently, the Web was mainly accessed by humans using web browsers and clicking on hyperlinks or submitting HTML forms. Clicking on a link is always a GET request while HTML forms only allow GET and POST methods~\cite{bug10671, issue195}. Recently, several web frameworks/libraries (like Ruby on Rails~\cite{railsform}, CakePHP~\cite{cakephpform}, Django~\cite{djangoform}, and .NET~\cite{dotnetform}) have started supporting RESTful web services through APIs. To support extra HTTP methods in browsers, these frameworks have used hidden HTML form fields as a workaround to convey the desired HTTP method to the server application. In such cases, the web server is unaware of the intended HTTP method because it receives the request as POST. A middleware between the web server and the application may override the HTTP method based on special hidden form field values. On one hand, this limitation is only with HTML and not Ajax requests. On the other hand, Ajax requests suffer from same-origin policy because support for Cross Origin Resource Sharing (CORS) is in the working draft of XMLHttpRequest~\cite{xmlhttp}. While modern web browsers have recently started supporting cross-origin Ajax requests~\cite{html5cors}, this feature is not available in old browsers.

Unavailability of the servers is another factor that affects the communication. Because of the stateless and synchronous nature of HTTP, a client must wait for the server to be available to perform the task and respond to the request. By introducing HTTP Mailbox as another layer of indirection, we can address these issues.

\section{Background}
\label{background}

\subsection{REST}

REpresentational State Transfer (REST)~\cite{fielding2000architectural, fielding2002principled} is a software architecture for large scale distributed system which has emerged as the preeminent design pattern. It utilizes existing HTTP methods to generalize the interfaces of the web service by mapping resource actions like Create, Read, Update, and Delete (CRUD)~\cite{crud} to corresponding HTTP methods POST, GET, PUT, and DELETE respectively. Remote Procedure Call (RPC) on the other hand encourages application designers to define their own application specific methods. A typical implementation of RPC on the Web is Simple Object Access Protocol (SOAP)~\cite{soap} that allows querying available procedures and associates arguments on a remote server then a client can invoke those procedures remotely using XML as the medium of exchange.

If Bob's tasks server on \url{example.com} was REST compliant, then after completing the first task Alice could have made an HTTP PATCH request to the task URI to mark it done as Code~\ref{rest} (Lines: 1-6) and gotten the modified task resource in response as Code~\ref{rest} (Lines: 8-12). It is not mandatory to return an entity body in the response to a PATCH request but in our example, we will assume that server will send the updated resource in the response. Media types \texttt{text/task} and \texttt{text/task-patch} are not defined, these are used here for illustration purpose only.

\begin{example}[caption=RESTful Communication,label=rest]
\end{example}
\begin{Verbatim}
> PATCH /tasks/1 HTTP/1.1
> Host: example.com
> Content-Type: text/task-patch
> Content-Length: 11
> 
> Status=Done

< HTTP/1.1 200 OK
< Content-Type: text/task
< Content-Length: 28
< 
< (Done) [HIGH] Write a paper.
\end{Verbatim}

Unfortunately, many web services are not fully REST compliant. Hence, a PATCH request as in Code~\ref{rest} (or other methods like PUT or DELETE) may end up getting \texttt{501 Not Implemented} or other failure responses. For example, the default Apache~\cite{apache} web server setup returns \texttt{405 Method Not Allowed} in response to a PUT request. Another issue is if Bob's server is not available then Alice has to wait and keep sending the request periodically until the service comes back online and completes the request.

Table~\ref{tab:methods} lists common HTTP methods and their support in web browsers and LAMP\footnote[1]{Linux, Apache, MySQL, and PHP, Perl or Python.} servers. It shows that Apache web server requires extra configuration in order to support PUT, DELETE and PATCH methods. Also, pure HTML has no interface to issue these methods from the browser except by using Ajax requests. It also gives statistical distribution of support of various HTTP methods on live web sampled over 40,902 random live URIs from DMOZ URI. We have selected the only URIs that return 200 OK response on GET request out of 100,000 initial set of URIs. Then we issued OPTIONS request on those 40,902 live URIs and collected data about supported methods from the ``Alow'' header. Only 55\% of live URIs responded to the OPTIONS request and only 1.16\% URIs returned all the methods listed in Table~\ref{tab:methods} in their ``Allow'' header. We did not check to see if the URIs respond to the methods returned in the ``Allow'' header. It shows the limited utilization of HTTP methods other than GET and POST on the web.

\begin{table}[h]
  \centering
  \caption{HTTP Method Support}
  \label{tab:methods}
  \begin{tabular}{p{1.2cm}|p{2.27cm} p{1.57cm} p{0.6cm} p{1.0cm}}
    \hline
    \textbf{{\color{black}Method}} & \textbf{{\color{black}LAMP}} & \textbf{{\color{black}HTML}} & \textbf{{\color{black}Ajax}} & \textbf{{\color{black}DMOZ}}\\
    \hline
    {\color{black}GET} & {\color{blue}Default Support} & {\color{blue}Link, Form} & {\color{blue}Yes} & {\color{blue}100\%}\\
    {\color{black}POST} & {\color{blue}Default Support} & {\color{blue}Form} & {\color{blue}Yes} & {\color{blue}40.3\%}\\
    {\color{black}PUT} & {\color{red}Extra Config.} & {\color{red}None} & {\color{blue}Yes} & {\color{red}1.7\%}\\
    {\color{black}DELETE} & {\color{red}Extra Config.} & {\color{red}None} & {\color{blue}Yes} & {\color{red}1.8\%}\\
    {\color{black}PATCH} & {\color{red}Extra Config.} & {\color{red}None} & {\color{blue}Yes} & {\color{red}1.3\%}\\
    \hline
  \end{tabular}
\end{table}

\subsection{Linda}

Linda~\cite{carriero1989lc} is a model based on generative communications~\cite{gencom} to facilitate distributed computing by sharing objects (e.g., data, computation requests and computation results) called tuples in a shared virtual memory called tuplespace. Processes query the tuplespace based on some criteria and perform a destructive or non-destructive read. Once the result of the process is ready, it is written back to the tuplespace where it can be picked up by another process.

Linda provides means for asynchronous (time-uncoupled) communication in which sender and recipient(s) do not need to meet in time. It also facilitates space-uncoupling as the sender and the recipient(s) do not need to know the identities of each other. This allows criteria based group messaging.

Linda has four basic operations or functions defined as:
\begin{itemize}
\item ``in'' -- a destructive read,
\item ``rd'' -- a non-destructive read,
\item ``out'' -- producing a tuple, and
\item ``eval'' -- creating a process to evaluate a tuple and producing a result tuple if applicable.
\end{itemize}

Now, assume that a client application on Alice's machine is communicating with Bob's task manager process via a shared tuplespace using the Linda model. To mark the first task completed, Alice's client may perform an \texttt{out} function to generate a tuple in the tuplespace for processing by Bob's service when available.

\begin{Verbatim}[numbers=none]
out("task", 1, "Done")
\end{Verbatim}

This means create a tuple for task with id 1 to mark it done. This tuple will remain in the tuplespace until Bob's service (or any other process) performs a destructive read using \texttt{in} function.

\begin{Verbatim}[numbers=none]
in("task", ?id, ?status)
\end{Verbatim}

This read query using the \texttt{in} function will match the Alice's tuple of ``task'', assign ``1'' to ``id'' and ``Done'' to ``status'', and remove it from the tuplespace.

Bob's service then can create a live/active tuple using the ``eval'' function to create a new process for marking the task with id 1 as done and update the tasks table to reflect the changes permanently. Bob's service may also wish to keep log of the changes.

\begin{Verbatim}[numbers=none]
eval("log", 1, changeStatus("Done"))
\end{Verbatim}

In this case, output of the live tuple will result in a passive tuple after the ``eval'' function is done, that can be stored in the tuplespace.

\begin{Verbatim}[numbers=none]
("log", 1, "Done")
\end{Verbatim}

This log tuple can be read using ``rd'' function several times without removing it from the tuplespace by Alice's client, Bob's server or any other entity that has access to the tuplespace. If Bob's service does not want to remove it from the tuplespace then a similar ``rd'' function can be used.

\begin{Verbatim}[numbers=none]
rd("log", ?id, ?status)
\end{Verbatim}

We took the simplicity of this model and considered implementing it on the Web scale to store and forward HTTP messages (requests and responses). Linda is a pre-web model mainly designed to work in a distributed system (not as large as the Web) where trusted processes share a common memory. Any process can write any tuple in the tuplespace independently and any process can destroy any tuple from the tuplespace. To implement it on the open Web as a distributed system, we must consider the scale of the Web and aspects of security and authenticity. Unlike a closed small distributed system, the Web is not trusted.

\section{Related Work}
\label{related}

\subsection{Relay HTTP}

The Relay HTTP draft specification~\cite{relayhttp} describes a way to overcome the CORS restriction posed by JavaScript in Ajax requests. A proxy service is built on the same domain to relay/replay HTTP requests between client and remote server. It uses \texttt{message/http} and \texttt{application/http} MIME types defined by~\cite{rfc2616} for tunneling HTTP traffic over HTTP. It requires additional setup on the Web server to host the proxy server.

Suppose that Alice wants to add a tasks block in her organization's website \url{example.org} while still utilizing the services of Bob's task manager hosted at \url{example.com}. She will fail to get data from or post data to the Bob's server using Ajax, because of the cross-origin restriction posed by JavaScript. Modern Web browsers which support CORS require additional headers from the server. But if she does not have control of Bob's server and if Bob's server does not already support CORS, she will not be able to get the tasks data from Bob's server. As a workaround, she may add an iframe in her website and embed Bob's tasks manager web page but she will not have control of the design of the embedded web page.

To overcome her client side restriction, she might set up a Relay HTTP proxy server under her organization's domain name (example.org) to delegate all cross origin requests to the proxy server to replay them on Bob's server and get the response as if it came from the same domain.

Using Relay HTTP, Alice makes a POST request which encapsulates the desired PATCH request as an entity to the proxy service hosted under her organization's domain hence avoiding any client side limitations as illustrated in Code~\ref{relay} (Lines: 1-11). Proxy service then replays the encapsulated \texttt{message/http} entity Code~\ref{relay} (Lines: 6-11) and forwards the response back to the client as Code~\ref{relay} (Lines: 13-17). But Relay HTTP still can not solve the server side limitations. Also, it is a synchronous system hence the client, the relay/proxy server, and the remote server must all meet in time.

\begin{example}[caption=Relay HTTP Communication,label=relay]
\end{example}
\begin{Verbatim}
> POST /proxy/example.com HTTP/1.1
> Host: example.org
> Content-Type: message/http
> Content-Length: 108
> 
> PATCH /tasks/1 HTTP/1.1
> Host: example.com
> Content-Type: text/task-patch
> Content-Length: 11
> 
> Status=Done

< HTTP/1.1 200 OK
< Content-Type: text/task
< Content-Length: 28
< 
< (Done) [HIGH] Write a paper.
\end{Verbatim}

\subsection{Bleeps}

Bleeps is a live messaging system that is inspired by tweets. It uses Push style communication~\cite{pushpull} to broadcast small messages using relay channels. Anyone can subscribe to one or more such channels to receive live message feeds. We explored Bleeps for the ResourceSync project~\cite{perspectrsync, techrsync}. Bleeps support hashtags and mentions for discovery and searching. It can be configured to support a variety of message formats for parsing message attributes easily using a language identifier. Messages are pushed to various channels for broadcasting which can be captured by consumer applications or other services. Bleeps are intended to be compact in length, preferably one-liners but structured enough to make the parsing easy according to the attached language descriptor. An example bleep looks like this:

\begin{Verbatim}[numbers=none]
from=alice to=http://example.com/tasks/1 change status #done @bob $task
\end{Verbatim}
%$

In this example, ``\$task'' at the end of the message is the language descriptor which defines the template for the message. Fields ``from'' and ``to'' can be used to query the message store. Similarly, ``\#done'' hashtag is there to help grouping the messages with the same status. Bob is being mentioned with the help of ``@bob'' which will cause the message to appear in Bob's stream. The remaining free text is the message which can also be a URL of a long message hosted elsewhere to keep the size of the message small. The message format is completely up to the attached language descriptor which can be defined by anyone.

\subsection{Enterprise Messaging System}

Apache Qpid~\cite{qpid} is an implementation of platform agnostic Advanced Message Queuing Protocol (AMQP)~\cite{amqp}. Java Message Service (JMS)~\cite{jms} defines reliable enterprise messaging standard. It is an integral part of J2EE platform. These messaging systems allow various modes of digital communication including point-to-point, peer-to-peer, pub-sub and other forms of individual and group messaging.

JMS is limiting as it is only for Java applications while Apache Qpid has servers (also called Message Brokers) written in C++ and Java, along with clients for C++, Java JMS, .Net, Python, and Ruby. However there is no easy way to interact with these messaging services using a web browser. There are plugins available for RabbitMQ~\cite{rabbitmq} (a message broker implementation for AMQP) that enable web communication (e.g., RabbitMQ-Web-Stomp~\cite{rabbitmqweb} that utilizes STOMP~\cite{stomp} protocol and WebSockets~\cite{rfc6455} to enable browser based interaction with RabbitMQ server).

\section{HTTP Mailbox Messaging}
\label{httpmail}

HTTP Mailbox messaging is a fusion of Linda style open access message storage and a traditional email system using HTTP as transport to embrace REST style asynchronous HTTP communication on the open Web. An HTTP Mailbox serves as a Linda style tuplespace for HTTP Messages.

\begin{figure}[ht]
\centering
\includegraphics[width=\linewidth]{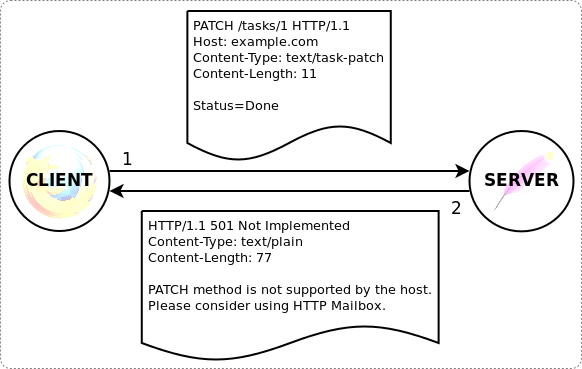}
\caption{Typical HTTP messaging scenario.}
\label{img:httpeg}
\end{figure}

\begin{figure*}[ht]
\centering
\includegraphics[width=\linewidth]{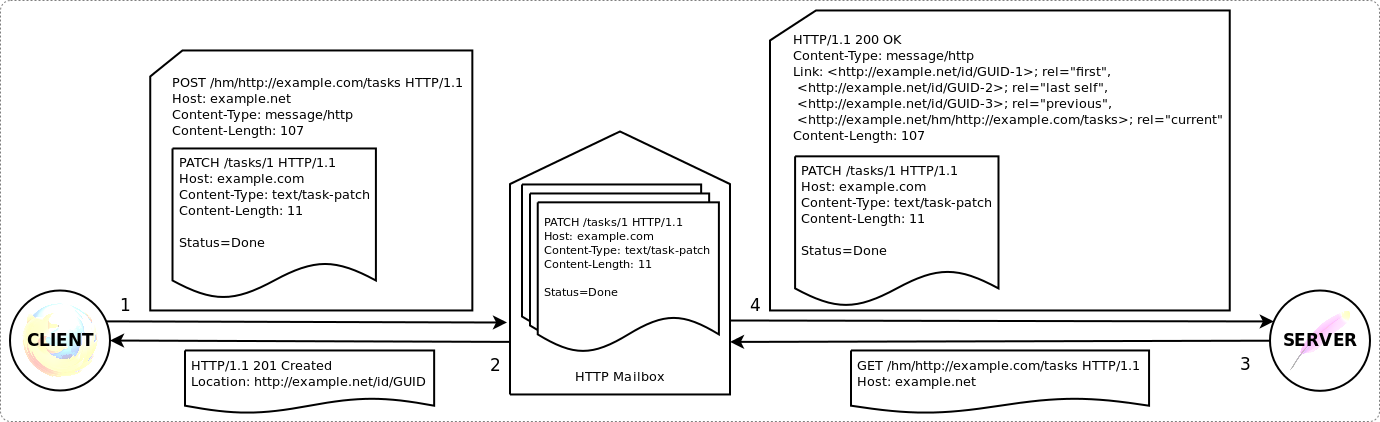}
\caption{HTTP Mailbox store and on demand delivery scenario.}
\label{img:hmeg}
\end{figure*}

In HTTP Mailbox messaging, HTTP requests are encapsulated inside another HTTP Message entity to form an envelope request. A client makes an HTTP POST Request to the HTTP Mailbox irrespective of the method of the encapsulated HTTP message. HTTP Mailbox then stores the encapsulated HTTP Request along with various message metadata in a persistent storage. Later, to retrieve those stored messages, a client makes an HTTP GET request to the HTTP Mailbox. Figure~\ref{img:httpeg} shows a typical HTTP Request and Response cycle. Figure~\ref{img:hmeg} illustrates how the same objective can be achieved using HTTP Mailbox while avoiding some of the issues of HTTP communication such as client or server side limitations and time coupling.

\subsection{Linda in HTTP}

One of the major advantages of HTTP Mailbox messaging is making HTTP communication asynchronous so that both the parties involved in the communication (typically known as ``client'' and ``server'') are time-uncoupled and do not need to meet in time for a successful HTTP communication (or a complete ``request'' and ``response'' cycle).

This asynchronous nature of communication is a good fit when a response from the recipient(s) is not necessary or not immediately needed. Hence we borrowed the ``store and forward'' model from Linda and transform it into a form that is suitable in HTTP environment on the scale of the Web.

A URI or any other identifier of the recipient(s) can be used to query messages from the distributed message store similar to the expressions used in ``rd'' and ``in'' functions of Linda to query the tuplespace.

The ``in'' function of Linda may be redefined as a soft-delete in the HTTP environment. We may not want to allow true deletion of messages because of the lack of trust on open Web and authentication challenges. Instead, flagging messages as deleted (and keeping a history of actions) may be a better choice because storage is not as limited as in case of pure Linda shared memory.

The ``rd'' function may exist without any modification and returning the message as many times and to as many clients as requested repeatedly. To facilitate additional functionalities, an access log may also be maintained.

The ``out'' function of Linda may refer to the action of preparing desired HTTP Request by a client and encapsulating it in another HTTP Request to send it to the message store.

The ``eval'' function of Linda may refer to the action of unpacking a stored HTTP request by a server, performing the desired task and writing the HTTP response in to the message store if necessary.

\subsection{HTTP Message}

``HTTP Request'' and ``HTTP Response''~\cite{rfc2616} both translate to a unified term ``HTTP Mailbox Message''. In order to complete the ``HTTP Request'' or ``HTTP Response'' transaction, both require a complete HTTP Mailbox messaging lifecycle.

From the HTTP Mailbox perspective, the restrictive terms ``client'' and ``server'' posed by ``HTTP Request'' (from client to server) and ``HTTP Response'' (from server to client) have disappeared and been replaced by the general terms ``sender'' and ``recipient''. But concepts of ``client'', ``server'', ``request'', and ``response'' continue to live inside the message body of the ``HTTP Mailbox Message''. To understand the differences at the encapsulated message level, ``HTTP Mailbox Message'' can further be subdivided into two categories, ``Indirect HTTP Request'' and ``Indirect HTTP Response'', in accordance with RFC 2616 ``HTTP Request'' and ``HTTP Response''.

\subsection{Indirect HTTP Request}

Suppose Alice is using an HTTP Mailbox service hosted on \url{example.net} to communicate with Bob's task manager service hosted on \url{example.com} from her organization's website hosted on \url{example.org} in a RESTful style.

Code~\ref{hmpost} (Lines:~7-12) is a typical HTTP PATCH request that she would send in order to mark the completed task done. Due to client or server side limitations (as discussed in section~\ref{intro}), a PATCH request may not be possible. Hence clients encapsulate the desired HTTP PATCH request in another HTTP POST request as illustrated in Code~\ref{hmpost} (Lines:~1-12). This POST request is made to HTTP Mailbox on a different \texttt{domain}, it has a different \texttt{path}, \texttt{Content-Type} and \texttt{Content-Length} as illustrated in Code~\ref{hmpost} (Lines:~1-5).

On a successful POSTing, HTTP Mailbox responds with a \texttt{201 Created} status code and provides a \texttt{Location} header with the URI of the resulting message as illustrated in Code~\ref{hmpost} (Lines:~14-16).

The request has not reached to Bob's server yet but now it is the responsibility of the HTTP Mailbox to deliver it when requested by Bob's server (in other words, when Bob's server pulls). Hence Alice's client is not blocked. In terms of Linda, thus far only the ``out'' function has been performed.

A client on behalf of \texttt{http://example.com/tasks} can then perform an HTTP GET request to the HTTP Mailbox as illustrated in Code~\ref{hmget} (Lines:~1-2) and get an HTTP response as illustrated in Code~\ref{hmget} (Lines:~4-21). This process is similar to the ``rd'' function of Linda.

\begin{example}[caption=POST HTTP Mailbox Request,label=hmpost]
\end{example}
\begin{Verbatim}
> POST /hm/http://example.com/tasks HTTP/1.1
> Host: example.net
> HM-Sender: http://example.org/alice
> Content-Type: message/http; msgtype: request
> Content-Length: 108
> 
> PATCH /tasks/1 HTTP/1.1
> Host: example.com
> Content-Type: text/task-patch
> Content-Length: 11
> 
> Status=Done

< HTTP/1.1 201 Created
< Location: http://example.net/hm/id/5ecb44e0
< Date: Thu, 20 Dec 2012 02:22:56 GMT
\end{Verbatim}

\begin{example}[caption=GET HTTP Mailbox Request,label=hmget]
\end{example}
\begin{Verbatim}
> GET /hm/http://example.com/tasks HTTP/1.1
> Host: example.net

< HTTP/1.1 200 OK
< Date: Thu, 20 Dec 2012 02:10:22 GMT
< Link: <http://example.net/hm/http://example.com/tasks>; rel="current",
<  <http://example.net/hm/id/aebed6e9>; rel="first",
<  <http://example.net/hm/id/5ecb44e0>; rel="last self",
<  <http://example.net/hm/id/85addc19>; rel="previous"
< Via: Sent by 127.0.0.1
<  on behalf of http://example.org/alice
<  delivered by http://example.net/
< Content-Type: message/http; msgtype: request
< Content-Length: 108
< 
< PATCH /tasks/1 HTTP/1.1
< Host: example.com
< Content-Type: text/task-patch
< Content-Length: 11
< 
< Status=Done
\end{Verbatim}

Two complete HTTP Request and HTTP Response cycles between a client and HTTP Mailbox, and a server and HTTP Mailbox respectively make one Indirect HTTP Request as illustrated in Code~\ref{hmpost} and Code~\ref{hmget} and shown in Figure~\ref{img:hmeg}

\subsection{Indirect HTTP Response}

After fetching messages from HTTP Mailbox and with the help of \texttt{Content-Type} and \texttt{Content-Length} headers as illustrated in Code~\ref{hmget} (Lines:~13-14), the server can parse the encapsulated HTTP PATCH request as illustrated in Code~\ref{hmget} (Lines:~16-21). The extracted HTTP PATCH Request can then be transformed (if necessary), executed on the task manager server and (if necessary,) a response may be sent to Alice using HTTP Mailbox as illustrated in Code \ref{resppost}. This process is similar to the ``eval'' function of Linda.

\begin{example}[caption=POST HTTP Mailbox Response,label=resppost]
\end{example}
\begin{Verbatim}
> POST /hm/http://example.org/alice HTTP/1.1
> Host: example.net
> HM-Sender: http://example.com/tasks
> Content-Type: message/http; msgtype: response
> Content-Length: 93
> 
> HTTP/1.1 200 OK
> Content-Type: text/plain
> Content-Length: 28
> 
> (Done) [HIGH] Write a paper.

< HTTP/1.1 201 Created
< Location: http://example.net/hm/id/32ab1ce2
< Date: Thu, 20 Dec 2012 02:31:12 GMT
\end{Verbatim}

\begin{example}[caption=GET HTTP Mailbox Response,label=respget]
\end{example}
\begin{Verbatim}
> GET /hm/http://example.org/alice HTTP/1.1
> Host: example.net

< HTTP/1.1 200 OK
< Date: Thu, 20 Dec 2012 02:42:03 GMT
< Link: <http://example.net/hm/http://example.org/alice>; rel="current",
<  <http://example.net/hm/id/26d1a9c2>; rel="first previous",
<  <http://example.net/hm/id/32ab1ce2>; rel="last self",
< Via: Sent by 127.0.0.2
<  on behalf of http://example.com/tasks
<  delivered by http://example.net/
< Content-Type: message/http; msgtype: response
< Content-Length: 93
< 
< HTTP/1.1 200 OK
< Content-Type: text/plain
< Content-Length: 28
< 
< (Done) [HIGH] Write a paper.
\end{Verbatim}

Later, Alice wants to check to see if her change was made, so she queries the HTTP Mailbox as illustrated in Code~\ref{respget}. If Bob's server updated Alice's task list and sent a response to the HTTP Mailbox then Bob's server response will be included in the HTTP Mailbox response to Alice's query as illustrated in Code~\ref{respget} (Lines: 15-19).

\subsection{Message Lifecycle}

\begin{figure}[!ht]
\centering
\includegraphics[width=\linewidth]{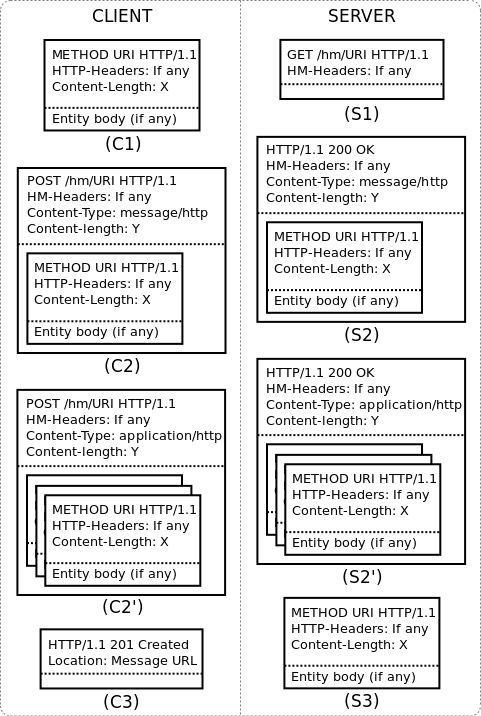}
\caption{HTTP Mailbox Lifecycle (top to bottom) on Client (Sender) and Server (Recipient) ends.}
\label{img:hm}
\end{figure}

A complete HTTP Mailbox messaging lifecycle consists of two phases, 1) Send, and 2) Retrieve. Each phase is further divided in two parts: request, and response. Each phase corresponds to one complete HTTP Request and Response cycle. Figure~\ref{img:hm} summarises the HTTP Mailbox process on both sender and recipient ends. C1) A generic HTTP Message (Request or Response), C2) An HTTP Message encapsulated in an HTTP POST Request to HTTP Mailbox (using \texttt{message/http} Media type), C2') Pipeline of one or more HTTP Message(s) encapsulated in an HTTP POST Request to HTTP Mailbox (using \texttt{application/http} Media type), C3) HTTP Response from HTTP Mailbox, S1) An HTTP GET Request to HTTP Mailbox, S2) An HTTP Message encapsulated in an HTTP Response from HTTP Mailbox (using \texttt{message/http} Media type), S2') Pipeline of one or more HTTP Message(s) encapsulated in an HTTP Response from HTTP Mailbox (using \texttt{application/http} Media type), and S3) A generic HTTP Message (Request or Response) extracted from the HTTP Mailbox Response.

\subsubsection{Send Request}

In the first phase of the HTTP Mailbox messaging a message is sent from the client (by or on behalf of a message sender) to the HTTP Mailbox server. This contains identifier of the recipient(s), some extra metadata and message body.

To send a message, an HTTP POST Request is made to the HTTP Mailbox server with recipients' identifier appended to the \texttt{HM-Base} of the mailbox as advertised by the HTTP Mailbox service on a well known URL (or root URL of the service), HTTP Mailbox service host as \texttt{Host} header, other extra metadata should go in HTTP headers (as discussed in section~\ref{header}). The entity must be a valid \texttt{message/http} or \texttt{application/http} request and appropriate \texttt{Content-Type} header must be present in the request headers (Figure~\ref{img:hm}~[C2, C2']).

\subsubsection{Send Response}

This is a feedback message from the HTTP Mailbox server to the message sender after receiving the ``send request'' message.

A status code \texttt{201 Created} will be returned along with the URI of the message in the \texttt{Location} header or an error code (e.g., \texttt{4xx/5xx}) in case of failure (Figure~\ref{img:hm}~[C3]).

A success response (status code \texttt{201 Created}) from the HTTP Mailbox server is a confirmation that the message has been stored and a promise that the message will be delivered whenever requested on behalf of the recipient(s).

\subsubsection{Retrieve Request}

The second phase of the HTTP Mailbox messaging begins with a message retrieval request from a client (by or on behalf of the recipient(s)). This request is made to the HTTP Mailbox server along with the identifier of the recipient(s) or direct URI of the message (if known), MIME type, and extra headers if necessary.

To retrieve the most recent message for a recipient, an HTTP GET request is made to the HTTP Mailbox server with recipients' identifier appended to the \texttt{HM-Base} of the mailbox as advertised by the HTTP Mailbox service, HTTP Mailbox service host as \texttt{Host} header, other extra metadata should go in HTTP headers (if necessary) while the entity must be empty. To retrieve an arbitrary message from the HTTP Mailbox server, an HTTP GET request must be made to the unique URI of the message (Figure~\ref{img:hm}~[S1]).

\subsubsection{Retrieve Response}

This is the final stage of the HTTP Mailbox messaging lifecycle. It is the response message from HTTP Mailbox server to the client when a ``retrieve request'' is made. It contains the message in the response body, MIME type and several other essential or optional headers in the header section of the response.

If the retrieval query was successful, a status code of \texttt{200 OK} should be returned along with \texttt{Via}, \texttt{Link}, \texttt{Memento-Datetime}, \texttt{Content-Type}, \texttt{Content-Length} and other optional headers (if necessary) followed by the message in the HTTP response body. The \texttt{Memento-Datetime}~\cite{memento} header contains the datetime when the message was first seen by the HTTP Mailbox. The \texttt{Link} header is used to provide navigational links to traverse the message chain back and forth (as discussed in section~\ref{chain}), identified by recipients' identifier. In case of success, the entity will be a valid \texttt{message/http} or \texttt{application/http} response and the appropriate \texttt{Content-Type} header will be present in the response headers. If the query does not match any messages or any other error occurred, an appropriate status code (like \texttt{4xx/5xx}) should be returned (Figure~\ref{img:hm}~[S2, S2']).

\subsection{HTTP Mailbox API}

One of the REST principles is Hypermedia as the Engine of Application State (HATEOAS)~\cite{fielding2002principled, alarcon2011hypermedia}. According to this a client needs no prior knowledge about how to interface with a RESTful service, except the generic understanding of relation types~\cite{linkrel} and MIME types~\cite{mediatype}. Client begins interaction with the service from a fixed URL and discovers future actions within the resource representations returned from the server. Clients should not rely on out-of-band information to interact with the RESTful service~\cite{resthyper}.

\subsubsection{HM-Request-Path}

An HTTP Mailbox service will advertise its base path (or base URL) for messaging called \texttt{HM-Base} (see appendix). This \texttt{HM-Base} will be used to construct request path (or request URI) at the time of sending or retrieving messages to or from HTTP Mailbox. In our examples, \texttt{HM-Base} is \texttt{/hm/}.

\texttt{HM-Request-Path} consists of two parts, \texttt{HM-Base} followed by recipients' identifier. Code~\ref{hmpost} and \ref{hmget} have \url{http://example.com/tasks} as recipient identifier in their first lines. Recipients' identifier can be a URI or any URL-encoded string token (like ``everyone'' for broadcasting or ``friends/alice'' for multicasting among Alice's friends).

Recipients' identifier may or may not match the path (or URI) in the \texttt{Request-Line} of the enclosed entity body. This is particularly important, because HTTP Mailbox is an on-demand message delivery service and it does not allow wild-card searching.

For example, if Alice's PUT request is to be sent to Bob's server to create a new resource at a non-existing URI, the \texttt{HM-Request-Path} may never be queried by Bob's server and the message will remain unread forever. In our examples, we have used \url{http://example.com/tasks} as recipient identifier while the enclosed entity has \url{http://example.com/tasks/1} as its URI.

\subsubsection{HM-Headers}
\label{header}

Apart from general HTTP-Headers, some headers are significant to the HTTP Mailbox. In a Send Request, the \texttt{HM-Sender} generic header should be sent, because the client sending the message may be sending it on behalf of someone else. On the other hand, in a Retrieve Response (in case of success) a \texttt{Via}~\cite{rfc2616} header is returned containing the identifier of the sender and the hostname or IP address of the sender client. A Retrieve Response (in case of success) must also return a \texttt{Link}~\cite{rfc5988} header containing \texttt{self}, \texttt{current}, \texttt{first}, \texttt{last}, \texttt{next}, and \texttt{previous} message URIs as applicable. HTTP Mailbox will also return a \texttt{Memento-Datetime} header to report the time when the enclosed message was first seen by the HTTP Mailbox.

In a Send Response a \texttt{Location} header containing the URI of newly sent message will be returned from HTTP Mailbox along with status code 201 if the message was successfully stored.

More HTTP Mailbox specific headers can be added later to extend the features of HTTP Mailbox like security, privacy, and message state.

\subsubsection{HM-Body}

All Send Requests and successful Retrieve Responses must contain \texttt{HM-Body} as entity body. \texttt{HM-Body} is a \texttt{Request} or \texttt{Response} \texttt{HTTP-Message} in one of the \texttt{message/http} (single) and \texttt{application/http} (pipeline) media types defined in~\cite{rfc2616}. Corresponding \texttt{Content-Type} header must be present in the \texttt{HM-Headers}. In other cases, there can be no entity body or any generic entity body with appropriate \texttt{Content-Type} header and \texttt{Status-Code}. In case of Retrieve Request, there must not be any entity body because it is an HTTP GET request.

\subsubsection{Message Chain}
\label{chain}

\begin{figure*}
\centering
\includegraphics[scale=0.5]{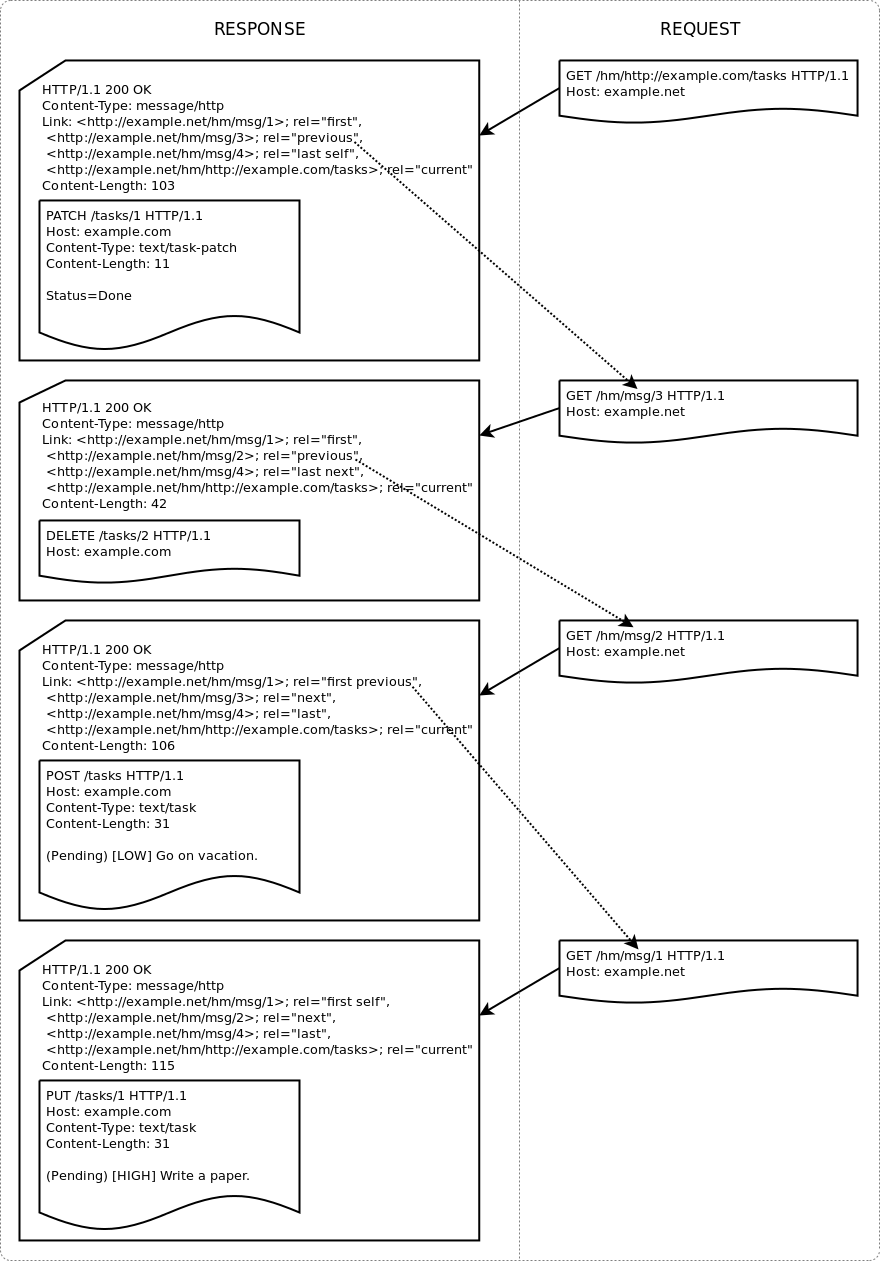}
\caption{Message Chain Retrieval.}
\label{img:hmchain}
\end{figure*}

The HTTP Mailbox query mechanism using \texttt{HM-Request-Path} allows the retrieval of the single ``most recent'' message sent to the corresponding recipient (if any). Every message also has a unique URI that can be used to Retrieve the message. By using the \texttt{Link} header of the response from HTTP Mailbox, an arbitrary number of messages or the entire message chain for the recipient(s) can be retrieved in either chronological or reverse chronological order, one message at a time. In a successful Retrieve Response, the \texttt{Link} header will contain the URI of the most recent message based on \texttt{HM-Request-Path} for the recipient as \texttt{rel=current}. It must also return unique URIs of \texttt{self}, \texttt{first}, and \texttt{last} messages with corresponding \texttt{rel} attributes. HTTP Mailbox will also return unique URIs of \texttt{previous} and \texttt{next} messages if present with corresponding \texttt{rel} attributes. Multiple \texttt{rel} attributes can be put together separated by a space if they point to the same URI. Absence of \texttt{next} relation and same values of \texttt{self} and \texttt{last}, both indicate the end of the message chain. Similarly, absence of \texttt{previous} relation and same values of \texttt{self} and \texttt{first}, both indicate the beginning of the message chain. Clients may use these indicators to detect either end of the message chain at the time of retrieval. Usually beginning of the message chain remains the same while end of the chain keeps changing over the time as more and more messages arrive for the same recipient. Figure~\ref{img:hmchain} shows a typical message chain retrieval scenario where the most recent message is retrieved first then it follows the \texttt{previous} link from the header until the first message in the chain is retrieved.

\subsubsection{Accessibility}

An HTTP Mailbox service should provide full CORS support so that restricted clients (like web browsers) can allow message sending and retrieval to and from the HTTP Mailbox while avoiding the JavaScript same-origin policy.

\section{Reference Implementation}

A reference implementation of an HTTP Mailbox server was written in Ruby~\cite{ruby} using the Sinatra~\cite{sinatra} Web framework running on a Thin~\cite{thin} Web server. Fluidinfo~\cite{fluidinfo} was used to store messages and other metadata associated with them and it was accessed using fluidinfo.rb~\cite{fluidinforb} Ruby library. A copy of this code can be found on GitHub~\cite{httpmailbox}.

\subsection{Benchmarking}

\begin{figure*}
\begin{center}
\subfigure[Send Message - POST]{\label{img:abpa}\includegraphics[scale=0.41]{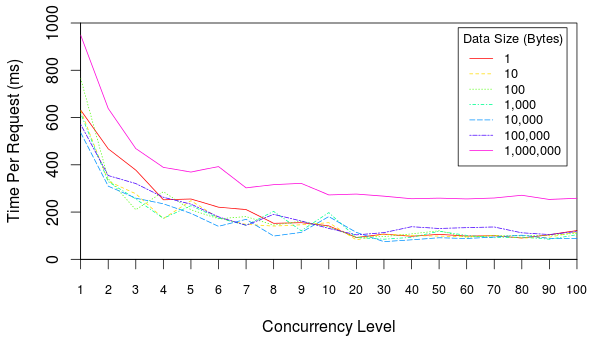}}
\subfigure[Retrieve Message - GET]{\label{img:abga}\includegraphics[scale=0.41]{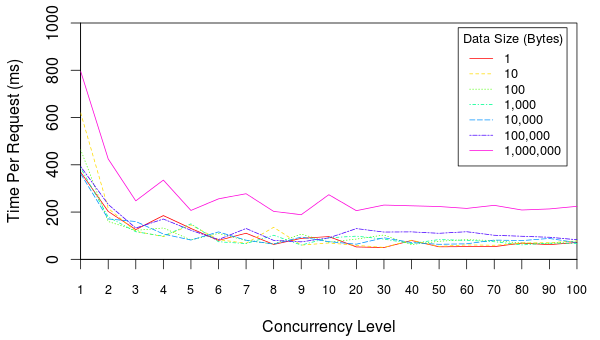}}
\caption{Stress Test Analysis of HTTP Mailbox using ApacheBench.}
\label{img:graph}
\end{center}
\end{figure*}

ApacheBench~\cite{ab} was used for stress testing of the reference implementation. We took digits of $\pi$ to generate payloads of various sizes ranging from 1 byte to 100,000,000 bytes ($\approx$~100~MB). Figure \ref{img:graph} shows the benchmark results of the HTTP Mailbox server (our reference implementation) on various concurrency levels and data sizes for Send and Retrieve requests respectively. The $Y$-axis shows the value of mean time per request (MTPR) in \textit{ms}. Each data point was generated by issuing total number of requests 10 times the concurrency level. The time taken by each request includes round trip time of the network time from benchmarking machine to HTTP Mailbox and message processing (which includes several HTTP connection between HTTP Mailbox and Fluidinfo server).

Graphs in Figure~\ref{img:abpa} and \ref{img:abga} show that the MTPR in both the cases decreases as concurrency increases. For smaller payloads (below 1~MB) the variation is not distinguishable, but when the payloads increase from 1~MB to 10~MB, MTPR roughly doubles. After analyzing data, we picked the 100~KB file and performed the stress testing up to a concurrency level of 1,000 and observed a gradual decrease in MTPR. With 1,000 concurrent requests, MTPR was 46 ms for POST requests and 34 ms for GET requests, while on concurrency level 100, these values were 118 ms and 83 ms respectively. ApacheBench socket did not allow more than a thousand open files for concurrent posting. Our implementation of HTTP Mailbox did not allow posting 10~MB (or larger) messages. We have also observed 0.0144\% (12/83,300) unexpected non-2xx responses in our benchmarking. While valid \texttt{404 Not Found} responses took 53 ms MTPR (values ranging from 200 to 20 ms depending upon the concurrency level).

\subsection{Preserve Me! Application}

A Web preservation application called ``Preserve Me!'' uses the services of the HTTP Mailbox heavily to fulfill its communication needs. This application is a JavaScript add-on utility that can be added in any web page. This will add a small ``Preserve Me!'' icon somewhere in the web page similar to several sharing icons (e.g., Tweet, Like, and +1).

\begin{example}[caption=Sending Add Friend Request,label=pmpost]
\end{example}
\begin{Verbatim}
> POST /hm/http://flickr.cs.odu.edu/rems/flickr-adittel-8162004738.xml HTTP/1.1
> Host: hm.cs.odu.edu
> Content-Type: message/http
> Memento-Datetime: Thu, 13 Dec 2012 05:15:55 GMT
> HM-Sender: http://arxiv.cs.odu.edu/rems/arxiv-0801-4807v1.xml
> Content-Length: 412
> 
> PATCH /rems/flickr-adittel-8162004738.xml HTTP/1.1
> Host: flickr.cs.odu.edu
> Content-Type: application/patch-ops-error+xml
> Content-Length: 270
> 
> <?xml version="1.0" encoding="UTF-8"?>
> <diff>
>   <add sel="entry">
>     <link rel="http://wsdl.cs.odu.edu/uswdo/terms/friend"
>       href="http://arxiv.cs.odu.edu/rems/arxiv-0801-4807v1.xml"
>       title="Automatic Text Area Segmentation in Natural Images"/>
>   </add>
> <diff>

< HTTP/1.1 201 Created
< Server: HTTP Mailbox
< Location: http://hm.cs.odu.edu/hm/id/5ecb44e0-859c-403f-9184-65e3a086ea2b
\end{Verbatim}

\begin{figure*}
\centering
\includegraphics[scale=0.6]{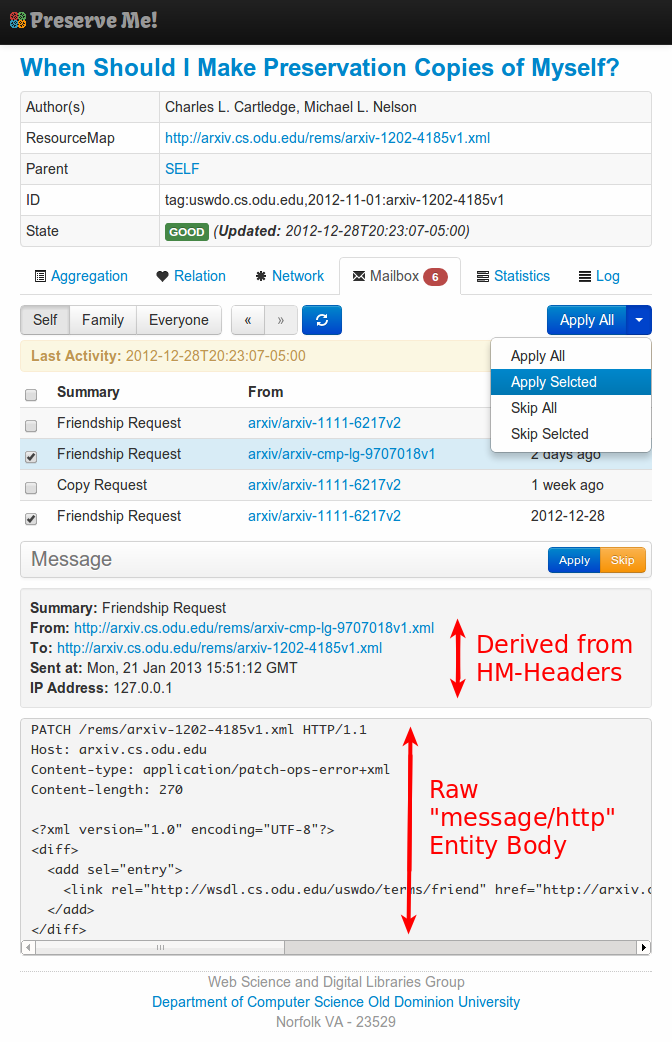}
\caption{``Preserve Me!'' Application window.}
\label{img:pm}
\end{figure*}

When that icon is clicked, it looks for one or more \texttt{link} tags with \texttt{rel=resourcemap} in the page. If the \texttt{href} attribute of those links points to a valid Atom ResourceMap~\cite{oreatom} files then it pops up a window as shown in Figure~\ref{img:pm}. This window gives insight into the ResourceMap and aggregated resources~\cite{oreatom} and also allows users to exchange messages with other ResourceMaps and connect them via ``family and friends'' relationships. These message exchanges and relationships allow human assisted preservation of aggregated resources and aids in their long term preservation.

\begin{example}[caption=Retrieving Add Friend Request,label=pmget]
\end{example}
\begin{Verbatim}
> GET /hm/http://flickr.cs.odu.edu/rems/flickr-adittel-8162004738.xml HTTP/1.1
> Host: hm.cs.odu.edu
> Content-Type: message/http

< HTTP/1.1 200 OK
< Server: HTTP Mailbox
< Content-Type: message/http
< Date: Thu, 13 Dec 2012 15:34:24 GMT
< Memento-Datetime: Thu, 13 Dec 2012 05:15:55 GMT
< Via: sent by 68.225.179.9
<  on behalf of http://arxiv.cs.odu.edu/rems/arxiv-0801-4807v1.xml,
<  delivered by http://hm.cs.odu.edu/hm/
< Link: <http://hm.cs.odu.edu/hm/http://flickr.cs.odu.edu/rems/
    flickr-adittel-8162004738.xml>; rel="current",
<  <http://hm.cs.odu.edu/hm/id/aebed6e9-e8ac-4051-9970-cc87fde2a549>;
    rel="first",
<  <http://hm.cs.odu.edu/hm/id/5ecb44e0-859c-403f-9184-65e3a086ea2b>;
    rel="last self",
<  <http://hm.cs.odu.edu/hm/id/85addc19-9358-46c7-a836-74b5161b2986>;
    rel="previous"
< Content-Length: 412
< 
< PATCH /rems/flickr-adittel-8162004738.xml HTTP/1.1
< Host: flickr.cs.odu.edu
< Content-Type: application/patch-ops-error+xml
< Content-Length: 270
< 
< <?xml version="1.0" encoding="UTF-8"?>
< <diff>
<   <add sel="entry">
<     <link rel="http://wsdl.cs.odu.edu/uswdo/terms/friend"
<       href="http://arxiv.cs.odu.edu/rems/arxiv-0801-4807v1.xml"
<       title="Automatic Text Area Segmentation in Natural Images"/>
<   </add>
< <diff>
\end{Verbatim}

\begin{table*}[t]
  \centering
  \caption{Feature Comparison of Various Messaging Systems.}
  \label{tab:feature}
  \begin{tabular}{p{3.0cm}|p{1.8cm} p{1.8cm} p{1.5cm} p{1.8cm} p{1.8cm} p{1.5cm}}
    \hline
    \textbf{Feature} & \textbf{HTTP} & \textbf{Linda} & \textbf{Relay HTTP} & \textbf{Bleeps} & \textbf{AMQP} & \textbf{HTTP Mailbox}\\
    \hline
    {\color{black}Multicast} & {\color{red}No} & {\color{blue}Yes} & {\color{red}No} & {\color{blue}Yes} & {\color{blue}Yes} & {\color{blue}Yes}\\
    {\color{black}Non-Blocking} & {\color{red}No} & {\color{blue}Yes} & {\color{red}No} & {\color{blue}Yes} & {\color{blue}Yes} & {\color{blue}Yes}\\
    {\color{black}Reliability} & {\color{blue}Yes} & {\color{blue}Yes} & {\color{blue}Yes} & {\color{red}No} & {\color{blue}Yes} & {\color{blue}Yes}\\
    {\color{black}Scale} & {\color{blue}Web} & {\color{red}Small} & {\color{blue}Web} & {\color{blue}Web} & {\color{blue}Web} & {\color{blue}Web}\\
    {\color{black}Message Size} & {\color{blue}Any} & {\color{blue}Any} & {\color{blue}Any} & {\color{red}Short} & {\color{blue}Any} & {\color{blue}Any}\\
    \hline
  \end{tabular}
\end{table*}

The Preserve Me! application sends various messages including: friendship request, copy request, and copy service announcements. Code~\ref{pmpost} and \ref{pmget} illustrate a typical friendship request message completing its lifecycle using HTTP Mailbox service. An aggregation (collection of resources) represented by a ResourceMap goes through a process of unsupervised creation of Small World Networks~\cite{cartledge2009unsupervised, cartledge2010analysis} to select various other aggregations as friends. To complete the friendship it requires the other aggregation to add a \texttt{link} tag in its ResourceMap, pointing back to the requester. To make that change, it prepares an XML-Patch~\cite{rfc5261} file as illustrated in Code~\ref{pmpost} (Lines: 13-20) and sends an HTTP PATCH request as illustrated in Code~\ref{pmpost} (Lines: 8-20) to the chosen friend. Because of the client and server side limitations, this request might not be directly  possible so it wraps the HTTP PATCH Message in an HTTP POST Message and sends it to the HTTP Mailbox as illustrated in Code~\ref{pmpost} (Lines: 1-20).

When the ``Preserve Me!'' window of the receiving aggregation is opened, it checks its Mailbox as illustrated in Code~\ref{pmget}. It then shows the friendship request (and any other messages, if available) as shown in Figure~\ref{img:pm}. If that message is applied then the friendship link will be added in the ResourceMap of the receiving aggregation.

\section{Evaluation}

Table~\ref{tab:feature} gives a quick overview of various features among various communication systems discussed in sections~\ref{background}, \ref{related}, and \ref{httpmail}. AMQP and the HTTP Mailbox are the two overall winners over the set of features listed in the table. While AMQP is a general purpose enterprise communication system, it is not very friendly for web communication especially using web browsers. On the other hand, the HTTP Mailbox is primarily made with RESTful web communication in mind.

\subsection{Availability}

Suppose that a sender has to send HTTP requests to $R$ number of recipients where an immediate response from the recipients is not required but the sender has to make sure that every recipient will eventually get the message. At any given time, a random subset of total recipients are unreachable but every recipient is mostly available over a period of time $T$.

In the HTTP communication, it may take a period of time as long as $T$ to successfully communicate with all the recipients and the sender has to make frequent attempts over time T. On the other hand, in the HTTP Mailbox communication, sender can send the message(s) to the HTTP Mailbox whose availability is much higher (assumed to be highly reliable) than individual recipients. The responsibility of eventual delivery of messages is then off loaded to the HTTP Mailbox and sender can proceed without being blocked.

\subsection{Network Usage}

The total number of HTTP cycles $C$ (where a cycle is combination of HTTP Request and Response) required to send $M$ messages to a group of $R$ recipients, assuming that there is no transient failure (or equally probable in all cases):

HTTP messaging:
\begin{equation}
C = M * R
\end{equation}

HTTP Mailbox messaging (where recipients only make attempts after sender has successfully sent the message to HTTP Mailbox.)
\begin{equation}
C = M * (R + 1)
\end{equation}

If the $M$ messages are grouped in $N$ ($\leq M$) message pipelines using \texttt{application/http}~\cite{rfc2616} MIME type then the cost of HTTP Mailbox communication will reduce further while cost of HTTP communication will remain the same.
\begin{equation}
C = N * (R + 1)
\end{equation}

In the worst case, individual unicast messages will cost twice for HTTP Mailbox communication as compared to HTTP. For larger group messaging scenarios it will cost roughly the same as HTTP while message pipelining will drop the cost of communication by a factor derived from the ratio of number of messages to the number of message pipelines. For simplicity, we have ignored the communication cost introduced by Pull~\cite{pushpull} attempts made by recipients before a new message arrived in the HTTP Mailbox for them, which is likely to happen because recipients are unaware of the sender's state.

\subsection{Response Pagination}
\label{pagination}

The HTTP Mailbox server itself can handle various concurrent requests but the design of HTTP Mailbox restricts individual message recipients from accessing their messages concurrently. As discussed in section~\ref{chain}, a recipient can not access arbitrary messages unless the URIs of those messages are known. After fetching a message a recipient only has URIs of first, last, next, and previous messages in the chain as applicable.

In our reference implementation the round-trip time of a single GET request with usual payload is between 300 to 400 ms as shown in Figure~\ref{img:abga} at concurrency level 1. As a result, a recipient can fetch roughly 3 subsequent messages per second from the message chain.

This issue is only limited to message retrieval (or GET requests). To overcome this problem, HTTP Mailbox can paginate responses. Every page can have a configurable number of subsequent messages that can be batched together along with the links to first, last, previous, and next pages to navigate through the chain of pages. To limit the scope, we have deferred the API definition for response pagination as future work.

\section{Future Work}

We are considering specifications for batch message retrieval (discussed in section~\ref{pagination}) and adding more query mechanisms like retrieving messages after a given timestamp. For access control, security, privacy, integrity, and authenticity~\cite{kaufmanpcpw}, we are planning to use techniques like OAuth~\cite{rfc6749}, public key encryption~\cite{rfc3447}, and hashing~\cite{rfc5754, rfc1321}. Data storage services other than Fluidinfo should also be evaluated to compare robustness and response time in each case. We are also planning to add message access log feature in the HTTP Mailbox and evaluate how it affects the utility and performance of the system.

\section{Conclusions}

In an effort of preserving web objects, we needed a messaging system that can be used reliably on the scale of the Web. We explored various possibilities including Linda, HTTP, and Bleeps but none of them fit our needs. Hence we have developed a store and forward model of HTTP messaging called HTTP Mailbox that remains RESTful and provides asynchronous (non-blocking) message sending and on-demand message retrieval facility between sender and recipients. It also provides message pipelining and group messaging (broadcast and multicast) facilities that save network usage and time.

Based on our model, we have implemented an HTTP Mailbox and tested its robustness and performance. Benchmarking our reference implementation gave us very reliable and time efficient results even on high concurrency levels within a data size limit. Unexpected failure rate was as low as 0.0144\% over more than 83,000 send and retrieve requests in our benchmarking.

We have successfully removed the client and server side barriers in using full range of HTTP methods in REST style. We have utilized our implementation of the HTTP Mailbox in the ``Preserve Me!'' application. We have also made the code of our implementation available on GitHub~\cite{httpmailbox}.

\section{Acknowledgements}

This work was supported in part by the NSF, Project 370161.

\bibliographystyle{abbrv}
\bibliography{httpmailbox}

\appendix

\section{Enhanced BNF}

\begin{Verbatim}[numbers=none,frame=none,fontsize=\scriptsize]
HM-Request-Path   = HM-Base ( http_URL | token )
HM-Base           = absoluteURI | abs_path
HM-Body           = HTTP-message
Send-Request      = "POST" SP HM-Request-Path
                    SP HTTP-Version CRLF
                    *( HM-req-header CRLF ) CRLF
                    HM-Body
Send-Response     = Response
Retrieve-Request  = "GET" SP HM-Request-Path
                    SP HTTP-Version CRLF
                    *( HM-req-header CRLF ) CRLF
Retrieve-Response = Status-Line
                    *( HM-res-header CRLF ) CRLF
                    [ ( HM-Body | message-body ) ]
HM-Header         = HM-res-header | HM-req-header
HM-req-header     = ( Sender-header | general-header
                    | request-header | entity-header )
HM-res-header     = ( Via-header | Link
                    | general-header | Memento-Datetime
                    | response-header | entity-header )
Via-header        = "Via" ":" "sent by"
                    SP (IP | IPv6 | Host)
                    SP "on behalf of" SP absoluteURI
                    SP "delivered by" SP absoluteURI
Sender-header     = "HM-Sender" ":" absoluteURI
\end{Verbatim}

\texttt{Link} is defined in RFC 5988~\cite{rfc5988}, \texttt{Memento-Datetime} is defined in~\cite{memento}, \texttt{IP} is defined in RFC 791~\cite{rfc791}, \texttt{IPv6} is defined in RFC 2460~\cite{rfc2460}, and remaining terms are inherited from RFC 2616~\cite{rfc2616} unless defined here.

\balancecolumns

\end{document}